\documentclass[conference]{IEEEtran}
\usepackage{cite}

\ifCLASSINFOpdf
\else
\fi
\usepackage{graphicx}
\graphicspath{{figures/}}
\usepackage{color}
\usepackage{amsmath}

\usepackage{cite}

\newsavebox{\exsavebox}
\newcounter{excounter}
\newenvironment{example}
     {\begin{lrbox}{\exsavebox}\begin{minipage}{0.98\columnwidth}\refstepcounter{excounter}Example \theexcounter: }
     {\end{minipage}\end{lrbox}\begin{center}\fbox{\usebox{\exsavebox}}\end{center}}

\newcommand{\boldtitle}[1]{\vspace{5px}\noindent\textbf{#1}}
\newcommand{\mycomment}[2]{\textcolor{red}{#1: #2}}
\renewcommand{\mycomment}[2]{}
\newcommand{\tb}[1]{\mycomment{TB}{#1}}

\newcommand{\tuple}[1]{\langle #1\rangle}

\newcommand{\PK}{\mathit{PK}}
\newcommand{\SK}{\mathit{SK}}
\newcommand{\CH}{\mathit{CH}}
\newcommand{\Rev}{\mathit{Rev}}
\newcommand{\Sig}{\mathit{Sig}}
\newcommand{\auth}{\mathit{auth}}
\newcommand{\validity}{\mathit{validity}}
\newcommand{\name}{\mathit{name}}
\newcommand{\Block}{\mathit{block}}
\newcommand{\Root}{\mathit{root}}

\begin{document}

%
\title{Application of Public Ledgers to Revocation\\ in Distributed Access Control}

\author{\IEEEauthorblockN{Thanh Bui}
\IEEEauthorblockA{Aalto University\\
thanh.bui@aalto.fi}
\and
\IEEEauthorblockN{Tuomas Aura}
\IEEEauthorblockA{Aalto University\\
tuomas.aura@aalto.fi}}


%


\IEEEoverridecommandlockouts
\makeatletter\def\@IEEEpubidpullup{9\baselineskip}\makeatother
\IEEEpubid{\parbox{\columnwidth}{
}
\hspace{\columnsep}\makebox[\columnwidth]{}}

\maketitle

\begin{abstract}
There has recently been a flood of interest in potential new applications of blockchains, as well as proposals for more generic designs called public ledgers. Most of the novel proposals have been in the financial sector. However, the public ledger is an abstraction that solves several of the fundamental problems in the design of secure distributed systems: global time in the form of a strict linear order of past events, globally consistent and immutable view of the history, and enforcement of some application-specific safety properties. This paper investigates the applications of public ledgers to access control and, more specifically, to group management in distributed systems where entities are represented by their public keys and authorization is encoded into signed certificates. It is particularly difficult to handle negative information, such as revocation of certificates or group membership, in the distributed setting. The linear order of events and global consistency simplify these problems, but the enforcement of internal constraints in the ledger implementation often presents problems. We show that different types of revocation require slightly different properties from the ledger. We compare the requirements with Bitcoin, the best known blockchain, and describe an efficient ledger design for membership revocation that combines ideas from blockchains and from web-PKI monitoring. While we use certificate-based group-membership management as the case study, the same ideas can be applied more widely to rights revocation in distributed systems.


\end{abstract}


%








\section{Introduction}
\label{sec:intro}

Blockchains and similar public log structures provide a new way to publish information and achieve consistency in a distributed system. The goal of this paper is to apply these ideas to distributed access control and, in particular, to the propagation of negative information such as revocations. We use group membership management with certificates as the main case study.

Bitcoin \cite{nakamoto2008bitcoin} was created as a new distributed currency that would challenge the current monetary system, and it soon received various competitors. The cryptocurrencies are, however, increasingly recognized for their broader contribution to the design of distributed systems. They combine several previously theoretical ideas to a new way of building a secure distributed system. Transactions take place between public keys and, thus, can be communicated as signed messages. The problem of global consistency is solved by compressing the global history of transactions into a public log structure called \textit{blockchain} and, finally, into one cryptographic hash value, which can relatively easily be agreed on and communicated to everyone. Inclusion and fairness are guaranteed by an open peer-to-peer (P2P) network, a competitive mining process, and monetary incentives in the form of transaction fees. Moreover, double-spending prevention, which is a global constraint on the permissible transactions, is enforced by the P2P network. Bitcoin challenged some fundamental ideas about what is feasible in a global distributed system. First, all local transactions are communicated throughout the global network and they are available for anyone to download. Second, the transaction data is not encrypted or fully anonymous. Previously, such compromises of performance and privacy would have been thought unrealistic, or outrageous. It is interesting to ask what distributed systems will look like if we accept the compromises made in Bitcoin and attempt to use them for other purposes than a monetary system.  

Recently, many new applications have been proposed that build on (or leech from) the Bitcoin P2P network and blockchain (e.g.~Colored Coins, Mastercoin, Factom, Proof of Existence, Tierion), and ones that propose constructing their own blockchain (e.g.~Ethereum, Namecoin). The overall observation in these schemes is that the blockchain has many uses beyond cryptocurrency implementation. Some sources use the generic term \textit{public ledger}, which stores information about \textit{transactions}, instead of blockchain, in order to avoid making a reference to the specific cryptographic implementation. Depending on the context, it may be easier to think of a \textit{public log} of \textit{events} instead. We use this terminology interchangeably. 

Access control in distributed systems is known to be a difficult problem. The lack of consistency that arises from parallelism and asynchronous communication may create race conditions \cite{bishop96racecond} that can be exploited by a malicious party. In particular, it is difficult in a distributed setting to communicate reliably negative information, such as revocation of access rights, or to verify the non-existence of credentials or revocation. A key observation that motivates our work is that the double-spending prevention in Bitcoin is also a kind of negative permission: once the money has been spent, it cannot be spent again. Thus, we ask the following question: \textit{Can public ledgers be used to enforce secure membership revocation and other flavors of negative permissions} in a distributed setting, and can it be done efficiently? 

The focus of this paper is on \textit{distributed group membership management} in a \textit{key-oriented system}, where entities are represented by cryptographic key pairs and group membership is granted with signed certificates. We use group membership as the main case study partly 
because it provides more complex and challenging revocation scenarios than most access-control systems. The work was additionally motivated by the practical need for a group-based access control mechanism in a credential-sharing service. 

This paper makes the following contributions: We show how the public ledger, as an abstraction, can be used to solve a central problem of distributed access control, the consistent distribution of revocation information. We use certificate-based group membership management as the case study and analyze its problems thoroughly. As our main technical contribution, we present a ledger design that solves these problems. To demonstrate the feasibility and efficiency of the solution, we implemented a prototype of certificate-based group membership management that supports membership revocation. The solution makes use of data structures from the security literature, which have previously been used for auditing the web PKI. We analyze the suitability of the Bitcoin blockchain for the same tasks and point out several limitations in it. More generally, we hope to increase understanding of the opportunities that the public ledgers offer for the design of secure distributed systems.




\section{Background}
\label{sec:background}

This section introduces concepts and related literature from blockchains and distributed access control, as well as related solutions for web PKI monitoring.  

\subsection{Bitcoin, blockchain and public ledger}
\label{sec:block_chain}



Bitcoin \cite{nakamoto2008bitcoin} was designed as a cryptocurrency, but it has found several other uses. It is widely used as a timestamping service (e.g.~CommitCoin \cite{clark2012commitcoin}, Factom, Proof of Existence, Tierion). Documents and other messages can be notarized by publishing a hash of the data in a Bitcoin transaction, 
which can later be used to prove that the document existed at a specific time. Other applications of the blockchain technology include the Internet name service (Namecoin) and anonymous credentials \cite{garman2013decentralized}. Recently, there has been a rush to propose new financial applications from land registry to diamond trade and from energy markets to tax audits \cite{ukgovernment16ledger}. Many of these applications build on the concept of smart contracts (e.g.~Ethereum \cite{ethereum}). These are financial contracts enforced in deterministic program code that can be executed by anyone independently to verify the outcome. Seeing the potential of the blockchain technology, mainstream organizations like The Linux Foundation and Microsoft are trying to advance it by developing open-source blockchain platforms\footnote{https://www.hyperledger.org/}\textsuperscript{,}\footnote{https://azure.microsoft.com/en-us/blog/bletchley-blockchain/}. Somewhat closer to the goals of this paper is the use of the Bitcoin blockchain as a PGP key server, which includes the publication of revocations \cite{wilson2015pretty}.

The success of Bitcoin and blockchain technology is due to the \textit{availability} and \textit{consistency} of its data. The blockchain contains the entire global history of transactions that have been created since the beginning and is publicly accessible by everyone. Moreover, since each block contains a reference to the previous block and requires a proof-of-work which is computationally expensive to produce, it is impractical for individuals to modify a block once it has been included in the blockchain. These properties protect the system from double spending (i.e.~spending more money that one has) and modification of previous transaction records. However, the nature of the blockchain makes it an \textit{inefficient structure for data retrieval}. Without additional indexes or summaries of transactions, looking for a specific transaction would involve going through all the blocks in the chain.

\subsection{Distributed access control}
\label{sec:distributed_ac}

In distributed access control, access rights are granted to subjects in the form of cryptographically protected \textit{credentials}. They are similar to the capabilities in the capability list \cite{lampson1974protection} of centralized access control, but instead of being stored with a central entity, they are distributed to clients. The credentials can be created with either secret-key \cite{gong1989secure, bull1992towards} or public-key mechanisms \cite{ellison1999spki, blaze1999keynote, blaze1996decentralized}. Our focus is on \textit{key-oriented} systems, which are not entirely dissimilar from the cryptocurrencies. In them, entities are represented by their public keys and the access credentials are in the form of signed \textit{authorization certificates}. The key-oriented PKIs were an active research topic in the late 90s as an alternative to the X.509 name certificates. While the proposed new certificate formats have not been standardized, they have been influential in the understanding of distributed access control and its limitations. The central ideas were coined, for example, in SPKI/SDSI \cite{ellison99usablepki,ellison1999spki,rivest96sdsi}, which encodes not only access rights but also name spaces and group membership with certificates. However, while we present our results in terms of signed certificates, the observations of this paper about revocation with public ledgers can be generalized to other types of distributed access-right and group management.  




Unlike traditional certificates such as X.509 \cite{housley2008internet}, which bind keys to names, the authorization certificates bind the keys directly to attributes or access rights. 
The basic format of an authorization certificate is the following:
\begin{equation}
\label{eq:certificate}
\begin{gathered}
C = \tuple{\PK_\mathit{issuer},\PK_\mathit{subject},\auth,\validity,\Sig_\mathit{issuer}}
\end{gathered}
\end{equation}

The certificate is signed by the \textit{issuer}, who gives to the \textit{subject} an \textit{authorization}, i.e.~access rights that are often related to a specific object, for a \textit{validity} period. The issuer and subject are identified directly by their public keys. With these certificates, an entity can delegate some of its authority to another. As a result, the certificates may form a chain or even a complicated network that reflects the underlying relations between their issuers and subjects. When a subject requests access to a resource, it sends a certificate chain to the \textit{verifier}. The verifier then determines whether the subject is permitted to perform the requested access operation by recursively verifying each certificate in the chain. 




\boldtitle{Revocation.}
Revocation is a fundamental operation in access control, and it is usually communicated with signed messages, which are similar to certificates but convey negative information. In the key-oriented PKIs, the effect of revocation is to invalidate one certificate, which also makes invalid all the certificate chains in which it has been included.


The problem of revoking authorization certificates is that the subject can choose which information to present to the verifier when making an access request, and nothing prevents it from excluding the negative information that would cause it to be denied access. Therefore, the verifier should have some independent means to \textit{verify the non-existence of negative information}. In Section~\ref{sec:revocation_pki}, we will apply a public ledger to solving this dilemma.




As will become apparent from the examples of Section~\ref{sec:revocation_group}, \textit{group management} may require another type of revocation where only group leaders can add or expel members from groups, and where members added by a previous group leader should remain valid even when the leader is revoked. 

\subsection{Public logs for monitoring the web PKI}
\label{sec:background_monitoring_web_pki}

The widely publicized compromises of certifiers, such as Comodo and 
DigiNotar, have motivated quite a few recent proposals for monitoring the security of the web PKI with the help of public log servers. 

Certificate Transparency (CT) \cite{laurie2013rfc} by Laurie et al.\ suggests public logs of all web certificates to bring transparency to the CA operations. Web clients should accept certificates only after checking that they are in the log. It is expected that, with the log, clients and special auditors can detect suspicious behavior, such as multiple certificates issued to the same name. The log is structured as an \textit{append-only Merkle hash tree}, in which new records are added to the right of the tree. The specification does not take a firm stand on who will maintain the log. In a follow-up blog post \cite{laurie2012revocation}, the authors suggested a \textit{sparse Merkle tree} for storing revocations, which enables constant-size (but relatively large) proofs of existence and non-existence for the revocations. They also hinted about separate trees for the time and lexicographic order. Ryan \cite{ryan2014enhanced} took these ideas further but replaced the sparse tree with a shallower hash tree, where certificates sorted by the subject name are stored in the in-ordered tree nodes. Again, the ordering allows logarithmic-size proofs of existence and non-existence for certificates. Ryan also points out the importance of auditing consistency between the two trees. 

In the Accountable Key Infrastructure (AKI) \cite{kim2013accountable}, certificates are registered in an \textit{integrity log server} (ILS). It maintains an \textit{ordered Merkle tree}, where the data in the leaf nodes is sorted by the domain name, thus enabling compact proofs of existence and non-existence. Independent \textit{validators} audit the ILS by maintaining a copy of the log tree. The web clients compare the root hash of the ILS with those of the validators. Revoked certificates are removed from the tree, so that it always stores the current status. The trees form a hash chain to enable auditing of the history, which is not very different from blockchains. ARPKI \cite{basin2014arpki} is a redesign of AKI which provides more prudent security guarantees by combining multiple X.509 certificates from different CAs to one certificate, and by establishing quorum among a fixed group of $n$ global ILSs. PoliCert \cite{szalachowski2014policert} is another similar solution with focus on domain-specific certification policies. In PoliCert, one Merkle tree contains both the certificates and revocation information, ordered by a hash of the certificate $H(C)$. In yet another variant of these ideas, PKI Safety Net \cite{szalachowski2016pki} enables verification of both the issuing time order and the non-existence of records for a domain by maintaining two trees that are similar to those in CT and AKI, respectively. Independent \textit{monitors} maintain a copy of the trees and audit their consistency. Unlike the earlier proposals, PKI Safety Net also considers certificate chains explicitly and allows revocation of intermediate CAs. CONIKS \cite{melara2015coniks} uses a \textit{Merkle prefix tree} to construct directories of user certificates. This also enables efficient proofs of existence and non-existence. Each leaf of the tree stores records that are mapped to the index represented by the path from it to the root. 

One lesson from these papers is that it is sufficient for one entity to maintain the public log if there are others who audit its actions. Ryan~\cite{dtki2014yu} points out, though, that such authorities should not be allowed to form an oligopoly. Most of the proposals (with the exception of \cite{ryan2014enhanced} and \cite{melara2015coniks}) are focused on X.509 name certificates for the web PKI or users. In this paper, we consider the use of public logs in the wider scope of key-oriented PKIs and in distributed group management.  



\vspace*{1em}
\section{Characteristics of a public ledger}
\label{sec:public_ledger}

In this section, we will discuss the properties of a public ledger. Although well known, it is worth stating them clearly. The goal is to lift the discussion from the level of cryptographic mechanisms to abstract characteristics, and to present a list of orthogonal features (although some interdependency cannot be avoided) from which ledger implementers and application designers can choose. In the later sections, we will observe the significance of these characteristic in distributed access control. 

\boldtitle{Immutability.} The ledgers are append-only, i.e.\ monotonically growing collections of transactions, or close approximations of such. It must not be possible to edit the already entered information or to roll back the ledger to an earlier state. In practical implementations, there may be a windows of uncertainty before the contents stabilize, and it may be possible to clean out entries that are no longer relevant. 

\boldtitle{Global consistency.} 
The most prominent property of the public ledgers is consistency of the ledger contents. That is, \textit{different parties retrieving information from the ledger will not receive conflicting information}. Together with the immutability property, this means that different parties have the same view of the current state and history of the ledger contents, and these views never diverge.  

Consistency is a widely studied issue in distributed systems, typically in relation to distributed databases and quorum protocols. Many different variants of the property have been defined, and there has been controversy about where blockchains fit among them. We do not intend to take a stand here, but rather note that there may be subtle differences between the kinds of consistency offered by different implementations of the public ledger, and we should remain observant about them.

\boldtitle{Inclusiveness.} 
Another necessary property of the public ledger is that \textit{it eventually accepts all entries sent to it, except those that violate defined constraints}. This is one type of \textit{fairness}, but since that word is heavily overloaded, we mostly avoid using it. Inclusiveness is important for access control because revocation of access rights and other negative permissions must not be blocked by those who maintain the ledger. 

Inclusiveness can be achieved with the help of third parties: We first try to enter a new transaction into the ledger and then check that it was actually entered. This check is reliable thanks to the consistency property of the ledger. If the ledger fails to include the transaction, we turn to the third parties whose job is to audit the correct behavior of the ledger. The ledger cannot refuse their request to enter the transaction, except when it can prove that the new transaction would violate a defined constraint. The process is optimistic in the sense that the third parties need to be involved only if the ledger behaves in an unfair way.  

\boldtitle{Linear order.} 
The public ledger also defines a linear order for all transactions. Unlike consistency and inclusiveness, the linear order is not an obvious requirement. It just turns out to be very useful for two reasons. First, sequential consistency \cite{attiya1994sequential} is one of the strongest and best understood definitions of consistency, and the explicit linear order makes it easy to design the consistency mechanisms of the ledger. Second, the linear order makes it easier to define various constraints (see below) on the transactions and to decide which of multiple conflicting transactions should succeed or fail. Sometimes, however, partial order is sufficient if different interpretations of it do not create inconsistent views of the ledger or enable constraint violations.

\boldtitle{Constraints.} 
A public ledger may define some global constraints that must hold at all times. These are akin to \textit{global safety properties} \cite{lamport77safety} in parallel and distributed systems. The constrains can be defined as global dependencies, such as total sums or as the non-existence of a negative permission. 

In Bitcoin and similar cryptocurrencies, the only global constraint is the over-spending prevention: transactions that would result in a key spending more of the currency than it has received may not be entered into the ledger. This constraint is defined in terms of the linear order of transactions, so that no key is allowed to run into credit even temporarily.  

In Sections \ref{sec:revocation_pki} and \ref{sec:revocation_group}, we will see examples of other global constraints related to revocation and negative permissions. 

\boldtitle{Reconstructible history.} 
In Bitcoin, it is possible to download and verify all the transactions that have taken place globally since the beginning of time. This creates confidence in the cryptocurrency because it shows the apparent fairness of how the money has been issued. In practice, though, someone using Bitcoin only needs to go back a few blocks in the history. Only those who join the P2P network of miners need to reconstruct the full history.  While this ability to download all past transactions and to reconstruct the global history is often seen as an essential part of public ledgers, it also puts a heavy burden on the participants. As we will see in Section \ref{sec:revocation_pki}, sometimes storing just a subset of the global events in the ledger is sufficient. 

\boldtitle{Efficient verification and management.} 
An interesting design choice in Bitcoin is that it is focused on verifying past monetary transactions rather than the current state of the system, i.e.~who holds money. While those in the P2P network keep track of the current state, other parties cannot trust a single member or even small part of the P2P network to provide this information. Instead, the canonical way to verify that someone has money is to make a payment with it and check that the transaction has been entered into the blockchain. This mechanism is, in part, because Bitcoin is focused on the overspending detection, and checking that someone holds money would not prevent them double spending it. In part, it is due to the way the P2P network is rewarded: without the fee paid for a new transaction, there is no incentive to construct proofs of the system state. In other applications of a public ledger, we may be interested in querying the global state without changing it. In access control, we want to verify access rights, role assignment, or group membership without ``spending'' them. 

As explained earlier, access control requires checking of non-existence. The Bitcoin blockchain is not efficient in verifying non-existence. First, it is not designed for efficient search or compact proofs of non-existence. Second, as a hash tree structure, the blockchain is extremely unbalanced if we consider the links to the previous block as the left-most branch. Proving non-existence of a transaction in the history is an O($T \cdot \log(\rho)$) process where $T$ is the length of the history and $\rho$ is the transaction rate. This can be compared to O($\log(T \cdot \rho)$) in balanced trees. Security literature provides more efficient hash-based data structures that allow fast searching based on an index and also checking of non-existence \cite{merkle79thesis,melara2015coniks,kim2013accountable}.

The literature also provides optimized solutions for implementing a public log with a small number of trusted parties. A particularly important observation is that only one (untrusted) third party is needed for maintaining the ledger if there are many auditors that verify its actions \cite{kim2013accountable}. Essentially, the auditors will need to check that the history never changes, compare the latest hashes of the history with other auditors, and check that any specified constrains on the ledger have been enforced. 

Finally, it is tempting to sort the ledger data or to create indexes or summaries to increase the efficiency of verification. However, maintaining these optimizations cannot be left to the care of an untrusted third party because they could be tampered with to produce inconsistent views of the ledger. Thus, the sort order, index or summary needs the same kind of auditing as the ledger itself. In effect, redundancy in the ledger creates new constrains that must be enforced and audited. It is necessary to compare the efficiency gains in access-rights verification to this increased cost of auditing. In this paper, we did not find examples where the benefits of adding redundancy would clearly outweigh the cost of auditing.


\vspace*{1em}
\section{Revocation in a PKI and hierarchical groups}
\label{sec:revocation_pki}


In this section, we discuss the revocation of certificate chains in a PKI and apply the public ledger to the task. Hierarchical group management with certificates is used as the case study. The main goal of this section is to introduce the issue of revocation and the basic solution components in a relatively familiar setting. We will build on this basis in the more complicated group-management scenarios of Section \ref{sec:revocation_group}.

\subsection{Certificate chains and group management}

In a PKI, each user or other entity is represented by a public-key pair $[\PK,\SK]$. The public key $\PK$ is used to identify the entity, while its owner knows the private key $\SK$. When requesting access to a resource, the user or the owner of the entity presents to the verifier a certificate chain 
\begin{equation}
\label{eq:chain}
\begin{gathered}
\CH = C_1\ldots C_n, \;\;\textrm{where}  \\
C_i = \tuple{\PK_{i-1},\PK_i,\auth_i,\validity_i,\Sig_{i-1}} \\
\textrm{ for } i=1\ldots n. 
\end{gathered}
\end{equation}
The details of the chain verification vary between systems, but the process typically includes at least the following checks: (1) the certificates form a chain so that the issuer of the next certificate is always the subject of the previous one, and the signatures are cryptographically valid, (2) the root key $\PK_0$ is authorized by some externa means to issue the first certificate, (3) the issuer of each certificate after that is authorized by the chain above, (4) the subject of the last certificate is the key that signed the access request, and (5) none of the certificates has expired or been revoked. We refer to these checks, excluding revocation, as \textit{general checks}. In the web PKI based on X.509, the root key $\PK_0$ is a trusted CA, while in SPKI, the root key may be the owner of the requested resource. In the web PKI, the issuers are authorized if the chain above delegates the CA role. In key-oriented PKIs (e.g.~SPKI), the rights passed through the chain can be computed as the intersection $\cup\; \auth_i$. 


\begin{example}
Alice uploads files to a folder at an online server and authorizes Bob to access the folder. Bob delegates the access further to his portable computer. When a file needs to be printed, the computer delegates the right to read that specific file to a cloud-based print server. If Alice revokes Bob's right to access, she would want both Bob's computer and the print server to lose their access without her taking any additional steps.
\end{example}

The key-oriented PKIs can be used to implement the above scenario so that each entity is represented by a public key and each delegation step is encoded as a certificate. The certificates chains are typically maintained in a distributed manner, so that the issuer gives to the subject not only the new certificate but also the preceding chain. When a client, such as the computer or print server, requests access to the controlled resource, it presents the chain of certificates, which starts from Alice, to the server. Revocation cannot be enforced in the same distributed way, though, because the server needs some way of checking whether revocations exist. 

Another observation is that we can interpret the entities in the above example as a \textit{hierarchical group} where Alice is the owner and Bob a member that is able to add further members to the hierarchy under himself. The hierarchy is more obvious in the following scenario.

\begin{example}
An online data storage authorizes its users to upload data to specific databases. The users authorize Internet-of-Things gateways, which in turn authorize the sensors that connect to them. The sensors data is sent to the database with end-to-end integrity protection. The user can revoke a gateway, in which case all the sensors connected to it should also lose access. If the user is kicked out of the service altogether, then all of his gateways and sensor should go as well. 
\end{example}


There are some general principles about revocation in PKIs. First, most PKIs allow revocation of certificates but not of keys. Key revocation would be equivalent to revoking all certificates issued to or by the key. Second, \textit{if any one of the certificates in the chain is revoked (or expires), the chain as a whole becomes invalid}. However, if a new certificate is issued to replace the revoked one, the chain can be formed again by reusing the rest of the certificates. Third, the revocation must identify the revoked certificate by some unambiguous means, such as by the combination of the issuer and a certificate identifier (which is a serial number in X.509) or by a cryptographic thumbprint of the certificate. Fourth, typically only the issuer of a certificate can revoke it, and the revocation must be signed by the same issuer as the certificate. Alternatively, the certificate itself could identify the authorized revocation key. Finally, fresh information about the relevant revocations must be available to the verifier in such a way that existence or non-existence of revocation can be reliably and efficiently determined. 

The availability question has caused quite a bit of debate. In the web PKI, the verifier periodically downloads the certificate revocation list from an address provided in the certificate, and hard and soft limits may be set on how often the list needs to be updated. SPKI, on the other hand, specifies no revocation mechanism because the delivery of the negative information in a truly distributed system cannot be guaranteed. SPKI tasks the entity making an access request with providing all the evidence needed for the access decision \cite{ellison99usablepki}, and they certainly cannot be trusted to tell about revocation. Instead, SPKI suggests periodic refreshing of short-lived certificates. This is considered a purer solution that makes explicit the difficulties of distributed access control and avoids the problems of best-effort revocation. In this paper, we ask whether it is feasible to make the revocation information available in a public ledger to get around such issues. 

Although we write about chains (i.e.~paths) here, the same ideas generalize to sets of certificates that form trees or directed acyclic graphs, in which multiple chains may be used together to authorize the issuer of the following one. This is the case, for example, in SPKI/SDSI, where the subject of a delegation certificate may be a name, in which case another certificate or chain is needed for resolving the name into a public key. 

\subsection{Revocation with public ledger}

It is relatively straightforward to implement revocation with a public ledger for the PKI applications discussed above. Because of the immutability and global consistency of the ledger, a revocation event cannot be deleted, altered or hidden once it has been entered into the ledger. Since the revocation is signed by the issuer of the certificate (or by another key defined in the certificate itself), the verifier can easily validate the signed revocation. The only difficult requirement for the ledger is that there must be an efficient process for retrieving revocations and for proving their non-existence in the ledger. Below, we use the thumbprint of the certificate, i.e.~its cryptographic hash value, as the index by which revocation information is stored and retrieved.





%
\boldtitle{revokeCert}: The issuer $I$ of certificate $C$ revokes it as follows:
\begin{enumerate}
\item Create a revocation record containing a hash of $C$:
\begin{equation} 
\label{eq:revocation}
\Rev_C \;=\; \tuple{H(C), \textrm{``revoke''}, \Sig_I}. 
\end{equation}
\item \label{revokeCert_submit} Submit $\Rev_C$ into the ledger with the index $H(C)$.
\item \label{revokeCert_check} Check that the revocation has been entered into the ledger.
\item If the ledger does not respond or refuses to accept a valid revocation record, raise an alarm.
\end{enumerate}

Depending on the ledger implementation, there may be a delay between steps \ref{revokeCert_submit} and \ref{revokeCert_check} above. In that case, the server should give a receipt of the submission 
to avoid delaying tactics and disputes about whether the submission took place. Ultimately, however, there can always be a dispute about which party failed to complete the protocol. In such situations, the solution is to submit the revocation through one or more well-known \textit{auditors}. Availability of such a fallback method removes the incentive from the ledger to misbehave.

\boldtitle{verifyChain}: The verifier checks whether the chain of certificates CH is valid for an access request as follows: 
\begin{enumerate}
\item Perform the general checks on $\CH$.
\item Calculate $H(C_i)$ for each certificate in the chain and query the public ledger using this thumbprint as the index. The public ledger must return either a revocation record signed by the issuer of $C_i$ or a non-existence proof. The chain is valid only if a non-existence proof is received for all the certificates. 
\item If the ledger does not respond or the returned information cannot be verified cryptographically, e.g.~because it is malformed, raise an alarm.
\end{enumerate}

For this type of PKI, the ledger does not need to hold any positive information, i.e.~the certificates. It is sufficient to make only the negative information, i.e.~the revocations, available in the ledger.

An alternative to signing the revocation is to authenticate it by revealing a hash pre-image. For this, the issuer generates a fresh random number $X$ and includes its hash $H(X)$ in the certificate. It keeps $X$ secret and stores it safely for later use. The issuer can revoke the certificate by publishing $X$ in the following revocation record in the ledger:
\begin{equation} 
R_X \;=\; \tuple{H(X), \textrm{``revoke''}, X}. 
\end{equation}
This solution does not suffer from the same kind of replay attacks as many other authentication schemes based on hash pre-images or Lamport hashes because replays can only help in making $X$ public. If the revocation record is published with the index $H(X)$, it is possible to use of the same $X$ in multiple certificates and to revoke them as a group. A possible extension is to include multiple hashes in the certificate so that it belongs to more than one such group. 


\subsection{Ledger implementation with Bitcoin}

We first consider using the Bitcoin blockchain for certificate revocation. There are several possible ways to encode the revocation messages of a key-oriented PKI (or even in X.509) into the Bitcoin blockchain. Any such method will probably make some use of the OP\_RETURN output that can embed up to 80 bytes of arbitrary data to one Bitcoin transaction. The main design choice is in how much of the structure of the Bitcoin transaction history is reused. We have identified the following two basic designs:

\begin{enumerate}

\item A Bitcoin address and, optionally, a revocation id is included in the certificate. The certificate can be revoked by making a payment from that address. The payment transaction must have either the certificate thumbprint or the revocation id in its OP\_RETURN output. 

\item The entire certificate is encoded as a Bitcoin transaction: the payer is the issuer, payee is the subject, and the OP\_RETURN output encodes the authorization and expiry time. Revocation is a similar transaction with a revocation flag encoded in OP\_RETURN. In this solution, the delegation graph is isomorphic with a subgraph of the Bitcoin transaction history. The 80-byte limit on data in OP\_RETURN is quite restrictive, though.
\end{enumerate}

The problem with the above solutions is the limited querying capability of the Bitcoin blockchain. There is no secure way to know about current status of the certificates, especially if one does not ``run a full node'', i.e.~receive and verify all the transactions in the global P2P network. Furthermore, while there are online services that can be used to search the blockchain data, if they are used for determining absence of items in the blockchain, they will effectively become trusted third parties. Moreover, the 60-minute cooling period required after a Bitcoin transaction means that even the most knowledgeable nodes cannot know the exact current state of Bitcoin. Transactions might be rolled back or buffered for inclusion in a later block. Thus, we will next consider solutions that involve a third party --- although not a fully trusted one. 


\subsection{Ledger implementation by third party}

We will now turn to a slightly more centralized implementation of the public ledger based on ideas surveyed in Section \ref{sec:background_monitoring_web_pki}. 

It is an inherent assumption in Bitcoin and its kin that no single party can be relied to operate any critical part of the system. Thus, these systems are structured around an open P2P network where all computation and data storage is massively replicated. Yet, most of the world's information infrastructure is operated by individual companies that have a business interest in keeping them running. While there are frequent questions about the honesty and fairness of such operators, nobody seems to suspect their will to stay in business. For example, one might suspect the honesty of an online poker site (blockchain solutions to this have been suggested), but there is hardly any reason to doubt that the site operator wants the gaming to continue. Thus, it is reasonable to try to treat continuity and honesty separately.

\begin{figure}[ht]
	\centering
	\includegraphics[width=0.35\textwidth]{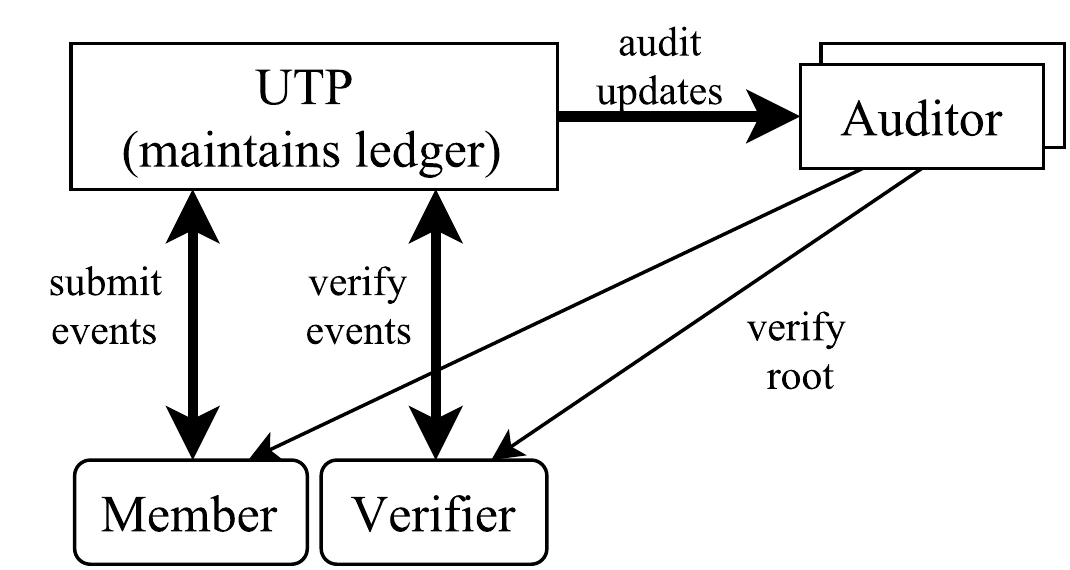}
	\caption{Ledger architecture with UTP and auditors}
	\label{fig:utp_architecture}
\end{figure}

\boldtitle{Ledger architecture.} The public ledger architecture that we propose to use for revocation is shown in Figure~\ref{fig:utp_architecture}. It is not based on a P2P network. Instead, we have one third party, called the \textit{untrusted third party} (UTP) to emphasize the lack of trust, and multiple trusted \textit{auditors} that monitor its honesty. This division of duties resembles for most part AKI \cite{kim2013accountable}. Our design goal is to leave heavy work to the UTP, while keeping the workload of the auditors and clients relatively small.  

We defer the full details of the ledger implementation until Section \ref{sec:revocation_group}. One possibility is to let the UTP store the lexicographically sorted revocations in the leaves of a Merkle hash tree, similar to AKI. Another possibility is a Merkle prefix tree like the one used in CONIKS. In the former case, the complexity of searching for an index $H(X)$ is $O(log(N))$ in which $N$ is the number of records. In the latter case it is $O(L)$ where $L$ is the length of the cryptographic hash, which should be long enough to avoid collisions. We build on the latter structure. Proofs of presence and absence in the tree will be covered below.

\section{Dynamic group membership revocation} 
\label{sec:revocation_group}

This section presents the main technical contribution of the paper. We discuss the problems of distributed, certificate-based group membership management in a dynamic setting where the group does not have a fixed hierarchical structure. We also describe a technical solution.


\subsection{Dynamic group membership and revocation}
\label{sec:membership}


We will extend the discussion of the previous sections to groups that have no strict hierarchical structure. This section introduces the new concepts through a series of example scenarios, which motivate the definitions that follow.

\begin{example}
An online course may have a group of admins and a group of users. The admins have the authority to add new admins and users and to remove existing ones. When Alice hands over the course to a new teacher Bob, she adds him to the admins. Bob then adds his teaching assistant Carol as an admin. Later, Carol cleans up the system settings and removes Alice, who no longer needs access, from the list of admins.
\end{example}

This is the kind of system that we aim to define and implement, but in a distributed setting where signed messages (i.e.~certificates and revocations) are the way to communicate and where negative is difficult to prove. The hierarchical PKI solution of Section \ref{sec:revocation_pki} will not quite work because Alice needs to be revoked by someone below her in the certificate chain, and because revoking Alice should not cause automatically the removal of Bob and Carol.

Since we are interested in systems of devices and services as well as users, we will use the words \textit{leaders} and \textit{members}, respectively, in place of admins and users. In this paper, these two \textit{roles} are assigned and revoked separately. In many applications, all leaders are members, and in some like the following example, or all members are leaders. Implementations can handle these situations efficiently by including multiple roles in a single  certificate or revocation message.

\begin{example}
David's personal devices form a group where his personal information is conveniently exchanged among the members. David can use any one of the devices to add a new device to the group or to remove one. When David buys a new phone, he first adds it to the group using his old phone. Later, when the new phone is fully operational, he sells the old one to his friend Erik and then remembers to remove it from his device group. 
\end{example}

We now start to see what the potential security issues are. Unlike the online course's user accounts, David's personal devices form a distributed system where the negative information of revoking a member may not be communicated immediately to all relevant parties. Moreover, conflicts may arise, as in the following.

\begin{example}
David's friend Erik is a l33t hacker and immediately recovers the device's secret credentials from its memory. He then tries to revoke David's new phone from the group in an attempt to take over the management of David's devices and data. Luckily, David remembered to revoke the old phone from the group in time, before Erik could remove the new phone from the group's leaders. 
\end{example}

Such race conditions are a natural part of distributed access control. While Erik in the story is the malicious party, the access control system is necessarily  agnostic about which device should win the race to revoke the other. Instead, our goal is to ensure that information about revocations is globally consistent and propagated reliably. No situation should arise where the current leadership of the group is ambiguous. This is where a public ledger can help.

We again take a key-oriented approach to group management. That is, users or devices are represented by key pairs and identified by their public keys. In our solution, any user can create a group by generating a key pair $[\PK_O,SK_O]$ (use of permanent personal keys is not recommended) and giving the group a name. The group is then identified by the combination of the public key $\PK_O$ and the name. 
\begin{equation}
G \;=\; G(\PK_O,\name)
\end{equation}

In a sense, all such groups exist without being explicitly created. Borrowing the notation of SDSI \cite{rivest96sdsi}, we could refer to this group as \textit{$\PK_O$'s name}. We take from SDSI the idea that names must be defined in the namespace of some public key but, because of revocation, the semantics of our groups will otherwise be quite different from SDSI. 

As in the hierarchical PKI, adding members will be represented by \textit{certificates}, which specify the role $R$ of the members in the group and are issued by some leader's key $\PK_L$ at a time $t$:
\begin{equation}
\label{eq:addmember}
C_U = \tuple{\PK_L,\PK_U,G,R,\textrm{``add''},t_L,Sig_L}_t, 
\end{equation}

These signed documents represent add-member \textit{events} that occurred at a specific time $t$, when they were issued. For now, let us assume that we have a global view of the system and know the time and order of the events, and that no two events occurred exactly at the same time. Of course, there are no reliable global clocks in a distributed system, and we cannot trust the signer's timestamp $t_L$ for ordering the events. Below, we will resolve this problem by defining $t$ as the order in which the certificates or revocations were added to the public ledger. 

Revoking members will be represented by signed \textit{revocations}, which are events similar to the certificates:
\begin{equation}
\label{eq:revmember}
Rev_U \;=\; \tuple{\PK_L,\PK_U,G,R,\textrm{``revoke''},t_L,\Sig_L}_t. 
\end{equation}

The main difference to Section \ref{sec:revocation_pki} is that these \textit{revocations apply to keys instead of certificates}. A revocation will invalidate a key's role in a group regardless of how many different certificates have previously been issued to assign it. Moreover, \textit{a group leader has the authority to revoke any members in its group} no matter which leader added them. Therefore, to verify a revocation, additional work may be needed to verify that the issuer of the revocation is a current leader. 

Finally, we get to \textit{define who is a group member or leader}. We do this by iterating through the events in the global time order. By definition, there is an initial event $e_0$ at time zero which makes the key $\PK_O$ leader of the group $G = G(\PK_O,\name)$ for all keys $\PK_O$ and all names. No other roles are initially assigned. Consider globally all the events related to the group $G$ and sort them to a global history by their timestamps $t$. This will produce a sequence of events $e_0 \ldots e_N$. An event $e_t$ is \textit{authorized} if its issuer $\PK_O$ is a leader in the group $G$ after the previous event $e_{t-1}$ in the global time order. The group membership after the event differs from the membership before the event as follows: If the event $e_i$ is authorized and it is a member certificate, then $\PK_U$ has new role $R$ in the group $G$ after the event. If the event is authorized and it is a revocation, then $\PK_U$ does not have role $R$ in the group $G$ after the event. These rules determine the group members and their roles through the global history of events. The current state of the system is the role assignment after the latest event in the global history.

\subsection{Membership revocation with public ledger}
\label{sec:revocation_group_ledger}

This section will consider how the public ledger, as an abstraction, can be used to implement secure revocation of dynamic group members. Since a revocation can be issued by any leader of the respective group, storing only the revocations in the ledger, as we did in Section~\ref{sec:revocation_pki}, is not sufficient. The verifier needs to also be able to check that the issuer of the revocation was a leader at the time. The challenge here is to choose which information is stored in the ledger and how to efficiently verify the issuer of any revocation.

We will start with a simple approach that is not efficient but illustrates the challenge. Naturally, we assume that the information stored in the ledger is immutable (i.e.~append-only) and the views that the ledger provides to different parties are consistent. Also, the ledger must accept all submitted valid events within a reasonable time. \textit{Beside the revocations, the clients also enter certificates into the ledger.} The ledger stores all events with an \textit{index} value, which is the cryptographic hash $H(G, R, \PK_U)$ of the group and role identifiers and the subject's public key. Anyone can query the ledger for the list of events with a given index value (the answer may be an empty list if there are no members). The ledger also provides a proof that this list is complete. This way, to check whether a user has a specific role in a specific group, the verifier can first query the certificates that have been issued to the user. It then queries the certificates that have been issued to the issuer of each of those certificates. This process is repeated until the verifier has found a valid certificate chain all the way back to the group's key $PK_O$. Although this is only a rough description of the process and other details need to be specified, it can be seen that checking the membership of a user with this solution might involve building the whole graph of relations between the group's members. Furthermore, if the ledger does not validate its input, malicious parties could insert fake certificates into the ledger to extend the graph so that the verification process becomes even more complicated.

To improve the efficiency of validity checks, we modify amend the above ledger design in several ways:

\begin{enumerate}
\item The ledger is used as a time-stamping service that gives all events (both member certificates and revocations) related to one group and role a global linear order. 
\item To add a revocation to the ledger, the issuer must prove with a certificate chain that it is a leader of the group and, thus, authorized to issue the event.
\item The ledger must check the validity of revocations before entering them into the ledger, and it must store the certificate chain that authorized the revocation. 
\end{enumerate}

In addition to item 2 above, we expect ledger implementations to also check the validity of member certificates before giving them a timestamp and entering them into the ledger. Such a check may be done to prevent denial-of-service attackers from filling the ledger with invalid events. However, this check is not needed for the correctness of membership management and we therefore omit it in the discussion below.

Item 3 in the list above is needed for efficiency reasons. More specifically, the ledger requires a new revocation $\Rev$ to be accompanied by a certificate chain $\CH_\textit{rev}$ that authorizes $\Rev$. The ledger validates $\CH_\textit{rev}$. This includes checking for the absence of revocations for all the certificates in $\CH_\textit{rev}$. The ledger then gives $\Rev$ a timestamp $t$, enters $\Rev$ into the ledger, and stores also $\CH_\textit{rev}$. Later, the ledger is able to present $\CH_\textit{rev}$ as an easily checkable proof that the revocation $\Rev$ was valid at time $t$. This will enable membership verification with computation and communication complexity $O(n)$, where $n$ is the length of the longest certificate chain in the ledger. 

As earlier, the member certificates are maintained by the members themselves. Each member $\PK_U$ of group $G  = G(\PK_O,\name)$ has a chain of certificates $\CH_U$ that starts from the key $\PK_O$. When a group leader adds a member to the group, it appends the new certificate to the chain and communicates the extended chain to the new member. When the member needs to prove its membership, e.g.~to make an access request or to submit an event to the ledger, it attaches this certificate chain to the request. The receiving party will then verify the chain and check the public ledger for revocations. 

The ledger gives each event a \textit{sequence number}, and this number is considered to be the global timestamp $t$. The ledger must assign these numbers sequentially (This is an internal constrain that the ledger implementation must enforce, as we will see later!). To detect outdated messages and replay attacks, the issuer of an event includes in the signed certificate or revocation the latest ledger sequence number that it knows $t_L$. The ledger requires $t_L$ in the event to be greater than the time of the last event recorded for the same index $H(G, R,\PK_U)$. This way, updates from leaders will be rejected if they are not based on the latest status of the key in the group and role. In addition to preventing accidental reordering of member certification and revocation events, this prevents attacks that intentionally delay events to cause reordering and, thus, change the outcome to something that was not intended by any of the issuers.  

The detailed processes are described below.

\boldtitle{addMember}: 
A group leader $\PK_{L}$ adds $\PK_{U}$ to role $R$ in the group $G$ as follows:
\begin{enumerate}
\item Issue a member certificate  
\[ C_U = \tuple{\PK_L,\PK_U,G,R,\textrm{``add''},t_L,\Sig_L}. \] 
\item Submit $C_U$ into the public ledger with index $H(G, R, \PK_U)$. Send to the ledger also a certificate chain $\CH$ that authorizes $\PK_L$ as a group leader. (The ledger is not required to check $\CH$ but may do so in order to refuse storing invalid member certificates.)
\item Check that $C_U$ has been given a timestamp $t > t_L$ and that it has been entered into the ledger.
\item If the ledger does not respond or refuses to accept $C_U$ without a valid reason, raise an alarm.
\end{enumerate}


Before the presenting the processes for revocation and verification, we define the following subprocess:

\boldtitle{checkChain} subprocess: 
A verifier checks that a certificate chain $\CH = C_1 \ldots C_n$ is valid and authorizes $\PK_U$ to role $R$ in the group $G = G(\PK_O,\name)$ as follows:

\begin{enumerate}
\item Perform the general checks on the chain. In the case of membership certificates, this means that (1) the certificates form a chain so that the issuer of the next certificate is always the subject of the previous one, and the signatures are valid, (2) the root key is $\PK_O$, (3) all the certificates delegate the leader role in $G$, except the last certificate, which delegates the role $R$, and (4) the subject of the last certificate is $\PK_U$.
\item \label{step:loopForRevocations} For each certificate $C_i$ in $\CH$, retrieve from the ledger its ledger timestamp $t_i$ and the list of events stored in the ledger with the index $H(G, ``leader",\PK_{i-1})$, where $\PK_{i-1}$ is the issuer of $C_i$. Receive also a proof of completeness of each list. From the list retrieved for each $i$, check that the leader role of the issuer $\PK_{i-1}$ has not been revoked between $t_{i-1}$ and $t_i$ (where $t_0=0$). For $\PK_U=\PK_n$, check that its role $R$ has not been revoked after $t_n$.
\item Return ``success'' or ``fail'' depending on whether all the above checks succeed.
If any one of the issuers $\PK_i$ has been revoked, return the revocation $\Rev$ in addition to the ``fail'' status.
\end{enumerate}

\boldtitle{revokeMember}: 
A group leader $\PK_L$ revokes role $R$ of $\PK_U$ in the group $G$ as follows:  

\begin{enumerate}
\item Issue a revocation: 
\[ Rev_U = \tuple{\PK_L,\PK_U,G,R,\textrm{``revoke''},t_L,Sig_L}. \] 
\item Submit $\Rev_U$ into the public ledger with index $H(G, R, \PK_U)$. Send to the ledger also a certificate chain $\CH_\textit{rev}$ that authorizes $\PK_L$ as a group leader. (The ledger must check this authorization and store the chain $\CH_\textit{rev}$. It can be stored in the ledger or in a secondary storage.)
\item Check that $\Rev_U$ has been given a timestamp $t$ and entered into the ledger.
\item If the ledger declines the revocation because $\CH_\textit{rev}$ is not valid, it must tell the reason. The only expected reason is a revocation $\Rev'$ in the ledger which makes the chain $\CH_\textit{rev}$ invalid and which the leader did not know about. \tb{What about a certificate being not available in the ledger? It leads to the question: does the subject need to verify the leader in addMember? } In that case, the ledger provides $\Rev'$. Retrieve from the ledger the chain $\CH'_\textit{rev}$ that authorized $\Rev'$, and check the correctness of the server's reason with the \textit{checkChain} process on $\CH'_\textit{rev}$. 
\item If the ledger does not respond or refuses to accept $\Rev_U$ without a valid reason, raise an alarm.
\end{enumerate}

\boldtitle{verifyMember}: The verifier checks whether the chain of certificates $\CH$ proves that $\PK_U$ has role $R$ in the group $G$ as follows:
\begin{enumerate}
\item \label{step:chainCheck} Check the chain $\CH$ with the \textit{checkChain} process. If \textit{checkChain} returns ``success'', then $\PK_U$ has role $R$ in $G$.
\item \label{step:revocationCheck} If \textit{checkChain} returns ``fail'' and includes revocation $\Rev$ as the reason, retrieve from the ledger the certificate chain $\CH_\textit{rev}$ that authorized $\Rev$. Check $\CH_\textit{rev}$ with the \textit{checkChain} process. If this second call to \textit{checkChain} returns ``success'', then $CH$ does \textit{not} give $\PK_U$ the role $R$ in $G$.
\item If the second call to \textit{checkChain} returns ``fail'', raise an alarm because the ledger is storing the revocation $\Rev$ without a valid authorization for it.
\item If the ledger does not respond or returns a syntactically or cryptographically invalid revocation record, raise an alarm. 
\end{enumerate}


\subsection{Security considerations}
\label{sec:revocation_security_considerations}

The above processes naturally should implement the membership semantics that we defined earlier in Section \ref{sec:revocation_group_ledger}.

\textbf{Proposition 1:} The processes described above enable the verifier to determine correctly whether a key is a member or leader of a group. 

We present informal reasoning to support Proposition 1:
In Section \ref{sec:membership}, we defined the semantics of group membership in the global history of member certification and revocation events. The certificate chain $\CH$ in step~\ref{step:chainCheck} of the \textit{verifyMember} process is a subset of the global history: one path of leader certificates from the initial event $e_0$ at time zero to the present time. The verifier first checks that, if we look at the subset alone, it would prove the membership. Now, the only events outside the subset that could change this outcome are revocations. Step \ref{step:loopForRevocations} of the \textit{checkChain} subprocess checks that no effectively timed revocations exist in the global history. Thus, if $\CH$ actually authorizes the membership of $\PK_U$ in $G$, the process returns ``success''. On the other hand, if there exists a revocation in the global history that invalidates $\CH$, then the ledger provides as evidence a such revocation $\Rev$ and the chain $\CH_\textit{rev}$ that authorized $\Rev$. The verifier calls \textit{checkChain} again to check that $\CH_\textit{rev}$ is valid. Again, the only events outside the subset of $\CH \cup \{ \Rev \} \cup \CH_\textit{rev}$ that could change the outcome are further revocations that invalidate $\CH_\textit{rev}$. The ledger must prove that no such revocations exist in the ledger. When the ledger does this, the verifier knows that $\CH$ does not authorize role $R$ in $G$ for $\PK_U$. The only reason why this might not happen is if the ledger actually contains a revocation that invalidates $\CH_\textit{rev}$, but that would be irrefutable evidence of the ledger's misbehavior because it stores the revocation $\Rev$ without having an authorizing certificate chain for it.  

While the reasoning above should satisfy us that the procedures implement the desired semantics, it does not mean that malicious entities cannot do anything harmful. For example, if a leader is compromised and acts fast, it can revoke other leaders and take over the group, or it may revoke everyone and thus destroy the group. 

One interesting case is a semi-malicious group leader who does not want to disrupt the group operation but wants to avoid being revoked from it. This could, for example, be the case in Examples 4 and 5 where Erik may want the phone to stay in David's device group, so that it continues to receive David's personal information. Such a semi-malicious leader, or a clique of them, could keep adding their own keys to the dynamic group at a fast rate, so that they become effectively irrevocable by other leaders. The group would still operate as normal in all other respects. In order to resolve this kind of situations, we allow a group leader to suspend all other leaders and temporarily reserve exclusive leadership in the group. It can do this by sending special suspend and resume events to the ledger: 
\begin{equation}
\label{eq:suspend_resume}
\begin{aligned}
Suspend_G \;&=\; \tuple{\PK_L,G,\textrm{``suspend''},t_L,\Sig_L}_t \\
Resume_G \;&=\; \tuple{\PK_L,G,\textrm{``resume''},t_L,\Sig_L}_t \textrm{.}\\
\end{aligned}
\end{equation}

These events are published in the ledger for transparency. Between the suspend and resume, $\PK_L$ is the only key allowed to issue new certificates or revocations to the group. It can thus analyze the membership and expel the unwanted leaders. Identifying the bad keys naturally requires processes that are outside the technical implementation. While the compromised leaders could maliciously suspend the group, that is no worse than other denial-of-service attacks that they can mount, such as revoking all members. What we achieve is preventing the semi-malicious attackers from hanging on for an unlimited time without doing anything explicitly malicious. 

The suspension is effectively a lock on the group in the ledger database, which makes is possible to execute complex processes as atomic operations. This feature can have further uses beyond the revocation of semi-malicious cliques. Naturally, it should only be used in exceptional circumstances because of the performance penalty that such locks have in a distributed system.


\subsection{Ledger implementation by a third party}
\label{sec:revocation_group_utp}


This section will show how the public ledger that was used as an abstract service in Section \ref{sec:revocation_group_ledger} can be implemented at a reasonable cost. We continue to use the architecture of Figure~\ref{fig:utp_architecture} with the untrusted third party UTP and independent auditors.

\boldtitle{Ledger data structure.} The UTP stores the ledger as a single \textit{Merkle prefix tree} in a way similar to CONIKS \cite{melara2015coniks}. A Merkle prefix tree is basically a binary tree where each path down the tree corresponds to a unique bit string $x$. Each bit in $x$ represents either a left or right turn on the way down. We denote by $V_x$ the node that corresponds to the $x$. If $V_x$ is a leaf node, it can be indexed by $x$ but also by a longer bit string that begins with $x$.

\begin{figure}[h]
	\centering
	\includegraphics[width=0.15\textwidth]{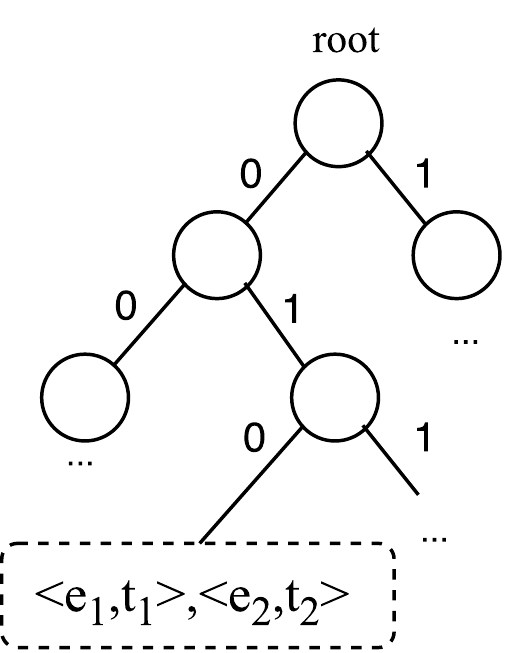}
	\caption{Merkle prefix tree with 2 entries stored in the leaf with index ``010" }
	\label{fig:log_tree}
\end{figure}

The events are stored in the leaves, which are indexed by the hash value $H(G, R,\PK_U)$. The events related to the same group $G$, role $R$ and subject key $\PK_U$ are bundled together into an append-only list, as seen in Figure~\ref{fig:log_tree}. The most recent record is at the end of the list. The list additionally stores a ledger-assigned \textit{sequence number} $t$, which we also call \textit{timestamp} because it is used in place of global time. 

We use SHA-256 for the hash function $H$ but only extend each branch of the tree down to the lowest level where no two input $\tuple{G, R, \PK_U}$ map to the same path. Thanks to the collision-freeness of the hash function, such a depth is always found. As more events are added to the tree, it slowly becomes deeper. 

The advantage of the single-tree structure for the ledger is that it is not necessary to audit consistency between two different trees. The timestamps, however, create similar internal constrains to the ledger data structure, and auditors are needed to enforce their sequential assignment. Note, however, that the UTP is responsible for linearizing the events and assigning the numbers and, hence, it is able to make minor changes to their order before entering them into the ledger. Minor reordering and gaps in the sequence numbers may also occur if multiple CPUs process updates in parallel and each is assigned a small range of timestamps at a time.

The UTP calculates a hash value $h_x$ for each node $V_x$ in the tree as follows. The hash of a \textit{leaf node} is a cumulative hash of the event list stored in the leaf node, including their sequence numbers, and the leaf node's full index (e.g.~256 bits as opposed to the 25-bit path to the leaf). The value of a non-leaf node is the hash of its two children:
\begin{equation} 
	h_x \;=\; H(h_{x0}, h_{x1})
\end{equation} 
However, if either the left or right branch of the tree does not continue, the child value is zero (either $h_x=H(0, h_{x1})$ or $h_x=H(h_{x0},0)$).

The hash value computed for the root of the tree, denoted by $h_\Root$, summarizes the whole data structure. Periodically (e.g.~once a minute), the UTP appends the latest value of $h_\Root$ into a hash chain along with the latest ledger and UTC timestamps:  
\begin{equation} 
  h^i_\Block = H( h^{i-1}_\Block, h_\Root, t_\textit{latest}, t_\textit{UTC} )
\end{equation} 
The values of this hash chain are signed and published to the auditors. The hash chain ensures that if the UTP ever forks or modifies the history, clients can detect its malicious behavior by comparing notes with each other or with the auditors. 




\begin{figure*}[t]
	\centering
	\includegraphics[width=0.85\textwidth]{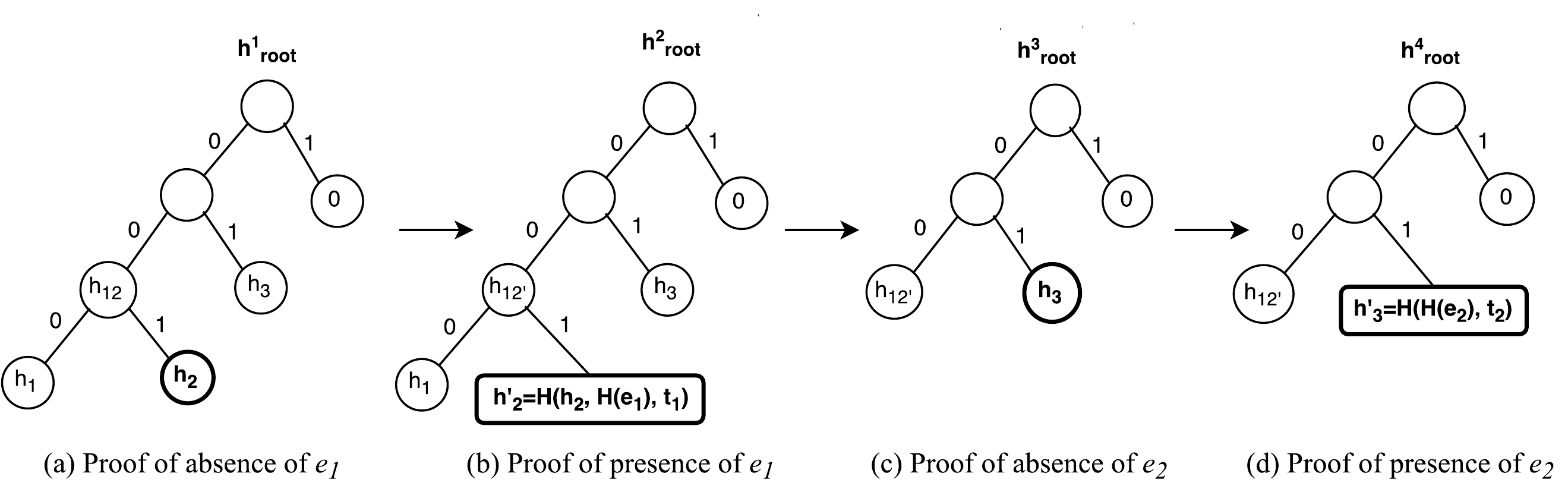}
	\caption{Example of how the root hashes form a chain when two entries, $e_1$ (with index ``001'') and $e_2$ (with index ``01''), are inserted to the tree between $h^1_\Block$ and $h^2_\Block$. The chain is valid if $h^1_\Root$ matches $h^1_\Block$, $h^2_\Root$ is equal to $h^3_\Root$ and $h^4_\Root$ matches $h^2_\Block$}
	\label{fig:audit_example}
\end{figure*}

\boldtitle{Ledger operations.}
The UTP needs to support operations that are required by the procedures described in Section~\ref{sec:revocation_group_ledger} and by the auditors. In particular, it must accept valid certificates and revocations from clients, allow the verifier to query presence of objects in the ledger by their index values, and send event data to the auditors for auditing. 

To check for the presence or absence of an event in the ledger, the UTP follows the path determined by the index $H(G, R,\PK_U)$ in the Merkle prefix tree. On the way down, it accumulates a \textit{proof} as a list of the hashes of the siblings of the path. That is, at any non-leaf node $h_x$ on the way down, if the next bit in the index is 0, the ledger appends $h_{x1}$ to the list, and if the next bit is 1, the ledger appends $h_{x0}$. If the search down the tree reaches a leaf node $V_x$ and the index stored in the leaf node matches the full index $H(G, R,\PK_U)$, the UTP appends the list of the hashes of events and timestamps stored in the leaf. This list is the \textit{proof of presence}. On the other hand, if the index stored in the leaf node does not match the full index, or if the search down the tree terminates at a non-existing branch, the so-far accumulated list of hash values becomes the \textit{proof of absence}. These are the proofs of completeness received by the client in Section \ref{sec:revocation_group_ledger}. 

When a client submits a revocation to the ledger in the \textit{removeMember} process, the UTP must verify it before adding it into the Merkle tree. For certificates submitted in the \textit{add\-Member} process, the verification is optional but probably will be done by most ledger implementations. For simplicity, we assume below that the ledger only contains verified events. The main part of the verification process is to check that the issuer of the new event is a group leader. For this membership check, the process is similar to \textit{check\-Chain} in the previous section. The only difference is that the UTP does not need to prove presence or absence as it trusts its own information. 

If the submitted event $e$ is valid, the UTP must immediately return a signed \textit{proof of delivery} (POD) before writing any data to the tree:
\begin{equation} 
	POD = \Sig_{UTP}( e, h^i_\Block, t_\textit{latest}, t_{UTC} )
\end{equation} 
The purpose of the latest block hashes and the ledger and UTC timestamps is to bind the receipt to the various notions of time in the system. The POD is, in effect, a promise by the UTP that it will include the certificate in the ledger as soon as it is technically possible. This receipt will be used as proof of UTP misbehavior if it fails to enter the event into the ledger. The user then has to wait until the next block update before it can verify that the event has been included. The user should not trust the UTP but periodically compares the UTP-presented block hashes with the ones published by the auditors. 

\boldtitle{Auditor operations.} 
The task of the auditors is to verify only the following:
\begin{enumerate}
\item The append-only property of the ledger, i.e.~that new events are added but old ones are not altered or deleted.
\item The timestamp $t$ for new events grows monotonically.
\end{enumerate}

There may be two kinds of auditors. The first kind maintains a copy of the entire ledger. It receives or downloads new events from the UTP and updates its copy of the ledger accordingly. It computes and publishes the block hashes $h^i_\Block$ for the benefit of the clients. It may also compare the latest block hash with the UTP's version as well as with other auditors. 

The second kind of auditor is what we have aimed for in the ledger design: the auditor receives from the UTP a stream of events with proofs of updates to the ledger. A \textit{proof of update} is simply a proof of absence followed by a proof of presence. Since the two proofs differ only for a very small part, sending the two does not take much more space than one. Furthermore, the UTP only needs to send the hashes of the events without any details since they are sufficient for the auditor to calculate the root hash. The proofs enable the auditor to compute the root hash before and after each update. The auditor does not need to save any ledger data. It simply checks that the root hashes form an unbroken sequence between two consecutive block hashes. It then publishes the verified block hashes as its view of the ledger. 

Figure~\ref{fig:audit_example} depicts examples of the tree updates with two new entries. Such updates and proofs of presence and absence are the main functions of the ledger. 



The auditor should store the latest UTP-signed block hash to prove misbehavior by UTP in case inconsistencies or mutations are detected in the history. Such errors must be published and will erode severely clients' trust in the UTP. Another task for the auditors is to help and act as a witness for any client whose valid events the UTP has refused and, if necessary, to submit them on the client's behalf. 

The auditors may be set up by major clients of the ledger, or they may be independent organizations. The second kind of auditor, i.e.~one that verifies the individual update proofs, can be implemented by multiple parties by taking turns. As long as the hash sequence between each two consecutive blocks is verified by at least one trusted auditor, the UTP cannot misbehave without being detected. It is a good idea for an auditor to audit multiple UTPs (i.e.~multiple ledgers), so that its own business is not threatened if it has to reveal misbehavior by the UTP.


\subsection{Combining two types of revocation}

In practical systems, we may encounter both types of groups: hierarchical groups where revocation invalidates the entire downward certificate chain (Section \ref{sec:revocation_pki}) and dynamic groups where leaders can revoke those above themselves in the delegation chain (this section). We can support the second type of revocation with the certificate $\Rev_C$ of formula \ref{eq:revocation}, which we reformat slightly to match the current system:
\label{eq:revrecursive}
\begin{equation}
Rev_C \;=\; \tuple{\PK_L,H(C),\textrm{``revoke''},t_L,\Sig_L}_t. 
\end{equation}

The certificate revocations will be stored in the ledger with the index $H(C)$. As with all revocations, the issuer presents an authorizing certificate chain $\CH_\mathit{rev}$ to the ledger, which stores it. When verifying a certificate chain $\CH$ (in \textit{checkChain}), the verifier queries the ledger for the index $H(C_i)$ on each certificate $C_i$ in the chain. If a revocation with authorizing chain is found for any $C_i$, the chain $CH$ is not considered valid.  

Because of their drastic effect, the certificate revocations should be used only in well-considered ways. Otherwise, a group leader could, for example, accidentally revoke itself and all other leaders, thus incapacitating the group. One way to use the certificate revocation is to tag a group as hierarchical, so that all revocation in it is recursive. Another solution is to tag a certificate as hierarchical and propagate this tag down the chain to newer certificates, which effectively creates a hierarchical subgroup. We expect that future experiments with such designs will clarify the best patterns.

\subsection{Ledger implementation with Bitcoin?}

It is possible to store arbitrary information in the Bitcoin blockchain and, thus, consider it published. However, we have not found a practical way of mapping the dynamic group membership management to Bitcoin so that it would support revocation. We could, quite elegantly, use Bitcoin for monotonically growing groups as follows. The Bitcoin addresses would represent users. The groups created by a user would be identified by the pair $\tuple{\textit{address, name}}$. The certificates would be small money transfers with the group name and member role encoded in the OP\_RETURN data. These transfers would be made from existing leaders to new leaders or members. Anyone could thus create a group and start issuing member and leader certificates. The only obvious way of revoking such membership, however, would be another transaction from the same leader who added the member, or possibly from the group creator. What we have not discovered is a mechanism by which an arbitrary group leader (from a different branch of the group) could revoke members added by the other leaders. One problem is that discovering such revocations requires extensive searching or traversal of the Bitcoin transaction graph. An even harder problem would be to check the absence of such revocations without keeping a full copy of the blockchain.



\section{Implementation and performance evaluation}
\label{sec:implementation}

\begin{table*}[t]
\begin{center}
\begin{tabular}{|l|r|r|r|p{0.2\textwidth}|}
\hline
\textbf{Solution} & \textbf{Memory} & \textbf{Bandwidth} & \textbf{CPU} & \textbf{Scalability} \\
\hline
Auditor maintains a copy of the ledger & 397 MB & 72 B/update & 19 607 updates/s & multi-threading possible \\
\hline
Auditor verifies individual update proofs & 288 B & 875 B/update & 10 526 updates/s & scales easily \\
\hline
\end{tabular}
\end{center}
\caption{Auditor resource use when the ledger has $10^6$ users and $10^7$ entries}
\label{table:auditor_cost}
\end{table*}

We originally designed our ledger mechanism for group-based access control in a credential management software where the ultimate goal was to share access codes, passwords and other information among groups of people or devices. In that system, data is shared through an online service operated by a company that does not want to become a trusted third party. (The details of how credentials are shared and accessed among the group are beyond the scope of this paper.) We implemented a prototype of the ledger and group membership system, including client, ledger and auditor, in order to demonstrate and evaluate its feasibility. The implementation supports both hierarchical and dynamic groups in the same ledger. The solution is generic and can be applied to group management in other distributed applications where there is an untrusted third party who is willing to maintain the ledger. 

\begin{figure}[tb]
	\centering
	\includegraphics[width=0.4\textwidth]{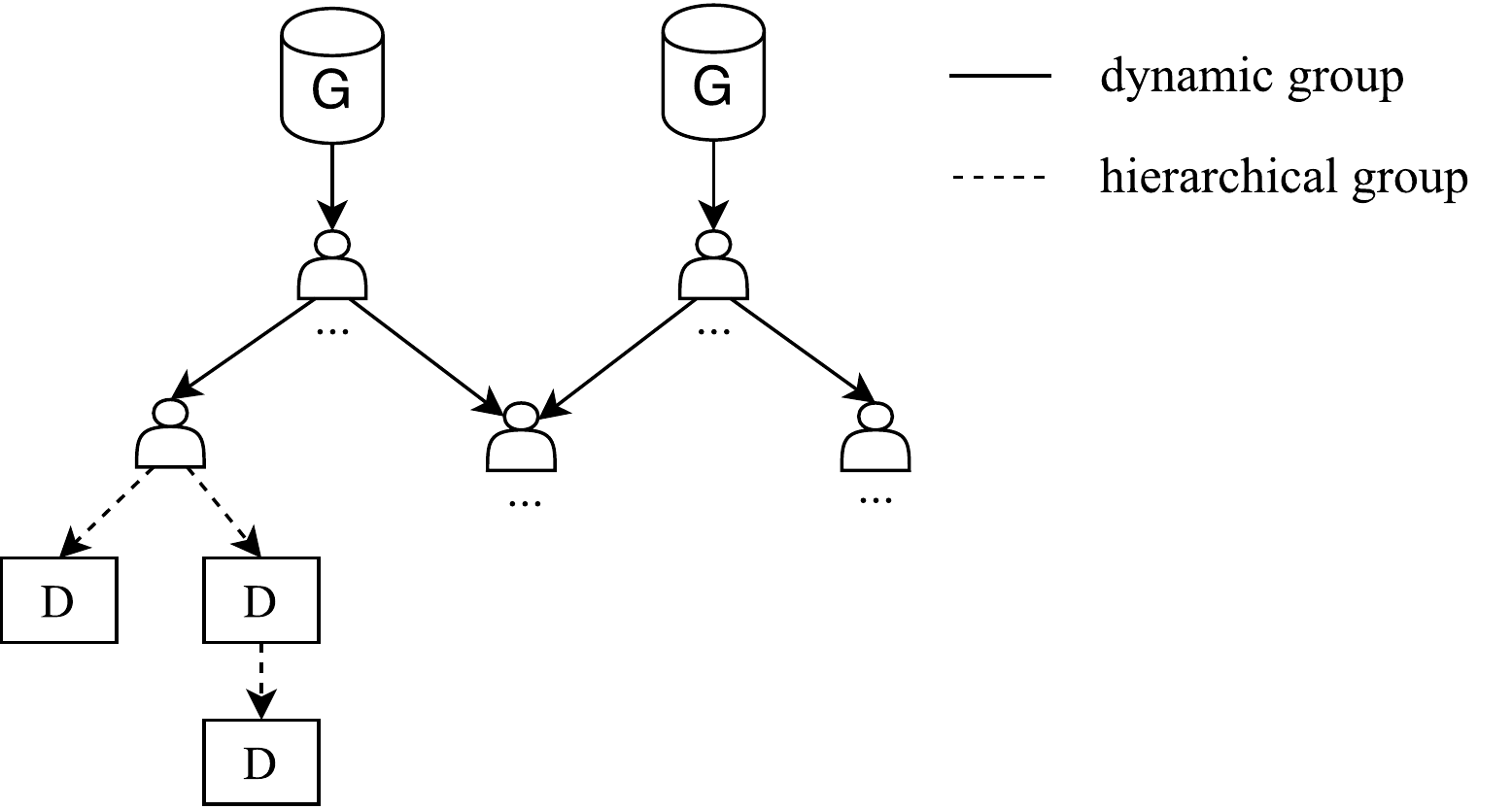}
	\caption{Two types of groups in the prototype}
	\label{fig:pm_architecture}
\end{figure}

Figure~\ref{fig:pm_architecture} illustrates how the two types of membership and revocation were combined in our functional test scenarios for the prototype: dynamic membership for groups of users, and hierarchical groups for one user's devices. Specifically, the system allows users to form groups, in which there are leaders and members. A user can form a group by creating a key pair, adding itself as a leader, deleting the group's key, and starting to add others to the group with its own key. The user can also delegate to his devices the membership in the groups to which he belongs. The devices can delegate the received rights further to other devices. The membership is represented by certificates, which together with revocations are published in the ledger as described in the earlier sections.  

Before running the performance measurements, we simulated $10^6$ users and devices that gradually formed groups until there were $10^7$ certificate and revocation entries in the ledger. The average path length in the Merkle prefix tree became 48. After this, we executed the performance measurements by verifying certificate chains and adding new certificates to the ledger. 

The prototype was written with Python (2.7.11) and the M2Crypto (0.25.1) cryptography library. We used SHA-256 as the hash function and RSA-2048 keys as the entity identities. We tested the prototype on a 3.4GHz Intel(R) Zeon(R) E3-1231 machine with 32GB of RAM. 


\boldtitle{Cost of adding certificates and revocations to the ledger.}
To insert a new certificate to the Merkle prefix tree, all the nodes on the path from the updated leaf to the root need to be updated. Our ledger server took, on average, 265 ns to insert a new entry to the tree and to collect the proof of update for the second-type auditors. This figure does not include signature verification. The average proof of update was less than 1 kB in size. 

When adding revocations, the ledger is required to verify the authorizing certificate chain. For adding certificates, the chain verification is optional but we chose to implement it. 

\boldtitle{Cost of verifying certificate chains.} 
We measured the average number of chain verifications that the server can handle per second when the length of the certificate chains was quite large ($L=50$). Since the verification does not involve updating the ledger, it is easy to parallelize. Our UTP server was able to verify on average 9385 certificate chains per second with an eight-core processor. Also, we expect the typical length of the chains to be shorter than 50, which will result in proportionally better performance.  

\boldtitle{Auditing cost.}
We compared two implementations of the auditor: one that maintains a copy of the ledger and one that verifies the proofs of update step by step without keeping any ledger data in memory. 

The memory use of the first auditor is naturally greater. The copy of the ledger requires $2*N$ hashes, which is $N \cdot 64$ bytes, for $N$ certificates and revocations. In contrast, the second auditor that verifies individual proofs does not require any significant storage space. 

The bandwidth use of the auditors depends on the transaction rate. The first auditor in our implementation needs to receive 72 bytes per update, which contain the index, the sequence number, and the hash of the entry. On the other hand, the second auditor without a copy of the ledger needs to receive the logarithmic-size proof of update, which was less than 1 kB in our experiments but grows logarithmically with the size of the ledger.


When running on a single CPU core, the first auditor was able to process about twice as many updates per second as the second one. This is expected because checking the update proof requires the auditor to verify both old and new hashes. The auditors' update rates were approximately $20000$ and $10000$ updates per second, respectively. 


With the first-type auditors, which keep a copy of the tree, the tree update process at the ledger and at the auditor can be parallelized as follows. The root node of the tree is a bottleneck because only one thread can be updating it at a time. This bottleneck can be avoided if the auditors maintain a copy of the prefix of the tree down to some depth $d$, e.g.~$d=10$, and the hashes in the prefix are recomputed only for the periodic block hashes. That way, the UTP can have up to $2^d$ processor working on different subtrees, without sharing any memory between them. Each processor's copy of the prefix will take an $2^{d+1} \cdot 32$ bytes, which is only 64 kB of memory for $d=10$.


\section{Discussion}
\label{sec:discussion}
This section describes the lessons that we have learned from this research and some possible extensions to the proposed ledger.  

\subsection{Lessons learned}

As we have seen, the public ledger is a suitable abstraction for implementations of distributed access control, including group-based ones, that require global consistency and enforcement of global constraints. This is the case, in particular, for revocation and negative permissions, which often cannot be checked without a global view of the system. Generic block-chain solutions enforce several fundamental security properties: immutability and global consistency of the event history, a linear order on all data entries, and fairness in the sense that all valid data is included. These properties are generic enough to be useful for most distributed access-control applications. We have, however, seen that different applications may have different requirements: they may need the public ledger to enforce special global constraints, and they may need different ways of indexing the ledger for data retrieval and proofs of presence and absence. This means that no one ledger architecture and implementation is sufficient for all purposes.

We also observed that some application constrains can be mapped to Bitcoin and its over-spending detection, while some do not map easily or at all. The main problem with Bitcoin, though, is the lack of support for querying the current state securely. The best view of the global state is held by the miners in the P2P network, but even they cannot be sure who holds money at the moment. Instead, the verifier can achieve certainty only by receiving the money: if the transaction is successful, it can be verified from the blockchain. But this is not the way access rights are queried. One would like to prove authorization without giving it away. 

It may be possible to map the concepts of distributed access control and group membership onto generic block-chain systems, such as Ethereum. Security literature, however, provides efficient public-log solutions based on (untrusted) third parties and independent auditors that form an alternative to the P2P network and competitive mining process of cryptocurrencies. They are also environmentally more sound as they waste less computing resources. Contrasting these more specialized public logs to blockchains helps in understanding the properties of both and may benefit research on both sides. In addition to avoiding the massive overhead and redundant work done in the P2P network, the alternative solutions fix the balance of the blockchain data structure, which can be seen as an extremely unbalanced hash tree. Our design and experiments show that such designs can work well for fairly generic distribute access-control tasks such as group management. 

One more technical lesson is that there are difficult trade-offs related to redundancy in the ledger data structure. On one hand, it can be ordered conveniently or enriched with additional index or summary information that helps application development; on the other hand, redundancy such as index trees, summary data, or even a monotonic numbering scheme, introduces security-critical internal constrains. It is necessary to check consistency between the ledger and the index or summary data. This checking may be too costly compared to the achieved efficiency gains. We note that revocation and negative permissions can be implemented relatively efficiently, and without redundancy in the data structure, when it is not necessary to check the mutual order of events but only their presence or absence. When the mutual order of all events needs to be stored, we show that a separate chronological tree is not needed. 

It is important to realize that even an ideal public ledger cannot solve all the problems of distributed access control. Because of asynchronous communication or temporary network partitioning, conflicting decisions by different authorities are a fact of life. By imposing a linear order on the events as they are recorded into the ledger, we can reach something akin to sequential consistency: the end result is in accordance with some --- arbitrarily chosen --- linear order of the submitted events. Together with fairness, i.e.~limit on the time anyone's data has to wait, this is probably the best we can expect. 

We also noted the need to avoid accidental or malicious reordering of events when it would change the outcome of the access control decision to something that was not wanted by any one of the distributed decision makers. Our solution was to reject entries to the ledger if they are not based on up-to-date state information. Such qualitative rules of freshness are preferable to timeouts because the round-trips in distributed systems can vary hugely from milliseconds to days, e.g., if a device is switched off or out of the wireless coverage area. 

\subsection{Possible extensions}

\boldtitle{Certificate validity times and temporary revocations.} 
The group-management solution presented in Section~\ref{sec:revocation_group} does not mention expiry times. Thus, the membership certificates remain valid until revoked. It would be tempting to include a UTC validity period in the certificates, similar to that in X.509. Handling such validity periods needs to be done with care, however. In X.509, the certificate verifier is only interested in whether the certificates are valid at the time of the chain verification, which it can do by comparing with a relatively accurate clock. In the dynamic group management scenario, on the other hand, the verifier needs to know the order of past events, which cannot be determined from the verifier's clock. Moreover, even very small differences in clocks or message propagation time could lead to different interpretations of whether a revocation is authorized or not. No matter how accurate the clocks are and how coarsely the expiry times are specified, a margin for such errors always remains if the expiration time is interpreted independently by different parties.  

The solution to this problem is to let the untrusted third party to decide on one consistent view of the order of all events, including the expiry of certificates. That is, the certificate expiry event (or beginning of the validity period) should be entered into the ledger when it occurs. This puts additional requirements on the auditors, though, because they must check that the UTP is entering the expiry events into the ledger at the right UTC times, within some specified error margin. 

This further calls our attention to the question whether a revocation can have a validity period. Such a temporary revocation, i.e.~suspension, of access rights can be useful, for example, when there is uncertainty about whether a lost device is compromised or in a safe place. Compared to revoking and re-issuing the certificate, the temporary revocation has the advantage that the group leader does not need to be active and online to revert the temporary revocation. We can implement temporary revocation by adding an optional UTC validity period to the signed revocation message. Just like with the certificates, the UTP should add the expiry events to the linear order of the ledger when they occur. 



\boldtitle{Role inheritance and RBAC support.}
So far, we have considered only two kinds of roles, \textit{leader} and \textit{member}, which are assigned independently. In practical systems, more roles could be defined, such as owner or guest. This calls into question the wisdom of having any fixed roles at all: perhaps the roles and their mutual relations should be defined for each application. Moreover, there could be some kind of inheritance between the roles to make their assignment easier. Indeed, the group-management solution presented in this paper is structured so that it can be extended with inheritance, role hierarchies and possibly other role-based access control \cite{sandhu1998role} features. We leave them as future work in order to focus on revocation, which is the main topic of this paper.

\section{Conclusion}
\label{sec:conclusion}

In the mad rush for new applications of blockchains, distributed access control seems to have been mostly an ignored area so far. It turns out that blockchains, and public ledgers as their generalization, are a nice abstraction for solving problems of distributed access control, and in particular those that arise from revocation and other negative permissions. Certificate-based group management, in particular, provides good test cases for understanding the requirements for public ledgers beyond their use in cryptocurrencies. Moreover, we observe that public-log constructions from security literature can teach useful lessons to those designing public ledgers. As the main result, we present a ledger-based design for distributed group-membership management and its experimental implementation.





\bibliographystyle{IEEEtranS}
\bibliography{references}
%

\end{document}